\documentclass[aps,prl,amsmath,amssymb,twocolumn]{revtex4}
\usepackage{graphicx}

\newcommand{\diff}[2]{\frac{d{#1}}{d{#2}}}

\newcommand{\torate}[1]{\overset{#1}{\to}}
\newcommand{\tofromrate}[2]{\underset{#2}{\overset{#1}{\rightleftharpoons}}}

\newcommand{\mol}[1]{\textsf{#1}}

\newcommand{\avg}[1]{\langle{#1}\rangle}
\newcommand{\m}[0]{{\rm min}}

\begin{document}
\date{\today}

\title{Reliability of frequency- and amplitude-decoding in gene regulation}

\author{Filipe Tostevin}
\affiliation{FOM Institute AMOLF, Science Park 104, 1098XE Amsterdam, Netherlands} 
\author{Wiet de Ronde}
\affiliation{FOM Institute AMOLF, Science Park 104, 1098XE Amsterdam, Netherlands} 
\author{Pieter Rein ten Wolde}
\affiliation{FOM Institute AMOLF, Science Park 104, 1098XE Amsterdam, Netherlands}

\begin{abstract}
In biochemical signaling, information is often encoded in oscillatory
signals.  However, the advantages of such a coding strategy over an
amplitude encoding scheme of constant signals remain unclear. Here we study
the dynamics of a simple model gene promoter in response to oscillating and
constant transcription factor signals. We find that in biologically-relevant
parameter regimes an oscillating input can produce a more constant protein
level than a constant input. Our results suggest that oscillating signals
may be used to minimize noise in gene regulation.
\end{abstract}


\maketitle

Cells are constantly exposed to a range of environmental stimuli to which
they must respond reliably. In recent years, it has become increasingly
clear that cells use complex encoding strategies to represent information
about the environment in the temporal dynamics of intracellular components
\cite{Behar10}. In particular, oscillatory or pulsatile signals are commonly
found in signaling and gene regulatory networks \cite{Paszek10}.  Perhaps
the best known example is the phenomenon of calcium oscillations
\cite{Berridge88}. Oscillatory dynamics have also been observed at the level
of gene regulation in nuclear localization of signaling proteins
\cite{Shankaran09} and transcription factors \cite{Nelson04,Jacquet03,Cai08}
and in transcription factor expression \cite{Lahav04}. Yet the advantages of
such a coding strategy for signal transmission remain unclear. 

A number of possible advantages for oscillatory signals have been suggested.
Oscillatory signals minimize the prolonged exposure to high levels of
calcium, which can be toxic for cells \cite{Trump95}. In systems with
cooperativity \cite{Gall00} an oscillating signal effectively reduces the
signal threshold for response activation. Pulsed signals also provide a way
of controlling the relative expression of different genes \cite{Cai08}.
However, these studies have ignored the impact of biochemical noise on the
reliability of signal transmission. Encoding of signals in protein
oscillations may play a direct role in ensuring accuracy in intracellular
signaling. 

Variability in the cellular response to an external signal will arise from
the temporal pattern of network activation and from inevitable biochemical
noise in the reactions making up the processing network, both of which will
depend on the coding strategy employed. A reliable response requires
minimization of such variability. Encoding of stimuli into oscillatory
signals can reduce the impact of noise in the input signal and during signal
propagation \cite{Rapp81}.  However, it remains unclear whether oscillatory
signals can also be {\em decoded} with a similar fidelity to constant
signals -- one might expect that the inherent variability of an oscillatory
signal would inevitably lead to a more variable response.  In this paper we
investigate whether oscillatory signals can be reliably decoded in a simple
model of gene regulation.  Surprisingly, we find that in
biologically-relevant parameter regimes an oscillating input can lead to a
more constant output protein level than a constant input. This effect arises
from differences in promoter state fluctuations, which it has recently been
shown can be a dominant source of noise in vivo \cite{Suter11}.

\begin{figure}[tb]
	\includegraphics{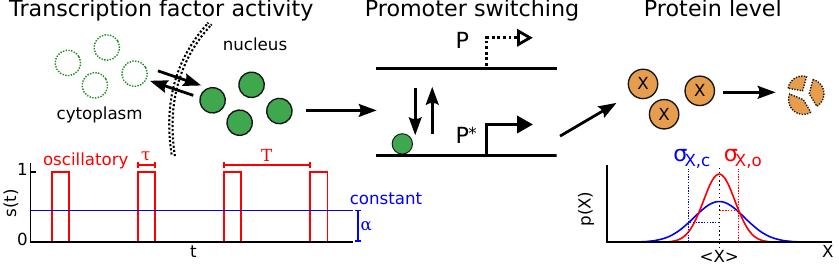}
	\caption{ \label{fig:intro}
		A gene promoter driven by constant and oscillating signals, $s(t)$. We 
		compare the variances of $X$ for the two input signals, 
		$\sigma^2_{X,{\rm	c}}$ and $\sigma^2_{X,{\rm o}}$, at the same mean level 
		$\avg{X}$.
	}
\end{figure}

We focus on the simple regulatory motif (Fig.~\ref{fig:intro}) of a single
gene promoter which can be in an active ($\mol{P}^*$) or inactive
($\mol{P}$) state.  The input to the system is the activity $s(t)$ of a
transcription factor which enhances activation of the promoter. A protein
$\mol{X}$, whose level $X(t)$ constitutes the network output, is transcribed
from the active promoter; proteins also spontaneously decay. The system
therefore consists of the reactions 
\begin{equation} \label{eq:promoter}
	\mol{P}\tofromrate{\kappa s(t)}{\lambda}\mol{P}^*,\ \ 
	\mol{P}^*\torate{\rho}\mol{P}^*+\mol{X},\ \ \mol{X}\torate{\mu}\emptyset.
\end{equation}
We will consider two forms for $s(t)$. In the first instance we take
$s(t)=\alpha$ to be constant. In the second, $s(t)$ is oscillatory,
reflecting the action of a periodic upstream signal or the inherently
oscillatory dynamics of the transcription factor. Since transcription factor
pulses often resemble distinct sharp peaks \cite{Jacquet03,Nelson04,Cai08}
we take $s(t)$ to be a binary switching process, with $s(t)=1$ for $nT\leq
t<nT+\tau$ ($n\in\mathbb{Z}$) and $s(t)=0$ otherwise (see
Fig.~\ref{fig:intro}). This form also has the advantage of making the
dynamics analytically tractable. We characterize the system response in
terms of the means $\avg{X}, \avg{P^*}$ and variances $\sigma_X^2,
\sigma_{P^*}^2$; for the oscillating input we define the stationary mean
$\avg{X}=T^{-1}\int_0^T\mathbb{E}\left[{X}(t)\right]dt$, where $\mathbb{E}
\left[\right]$ denotes averaging over network realizations with the same
input, and the stationary variance
$\sigma_X^2=T^{-1}\int_0^T\mathbb{E}\left[X(t)^2\right]dt-\avg{X}^2$.  These
are calculated from the chemical master equation \cite{VanKampen}, and
verified by stochastic simulations \cite{Gillespie76}. 

When driven by a constant signal the promoter simply undergoes random
switching with constant rates. Hence the probability of the promoter being
active at any time is $\avg{P^*}_{\rm
c}=\alpha\kappa/(\alpha\kappa+\lambda)$. For an oscillating input the
average activity,
\begin{equation*} \label{eq:promoter_fm}
	\avg{P^*}_{\rm o}=\frac{\kappa}{\kappa+\lambda}\left[\frac{\tau}{T}+
		\frac{\kappa(1-e^{-(\kappa+\lambda)\tau})(1-e^{-\lambda(T-\tau)})}
			{\lambda T(\kappa+\lambda)(1-e^{-\kappa\tau-\lambda T})}\right],
\end{equation*}
contains a term from the expected promoter activity when $s=1$ multiplied by
the fraction of time, $\tau/T$, for which $s(t)=1$, and a correction due to
the fact that the promoter response timescales when $s(t)=1$ and $s(t)=0$
($[\kappa+\lambda]^{-1}$ and $\lambda^{-1}$, respectively) differ. Since
promoter switching is a two-state process the variance in promoter activity
is determined by the mean, $\sigma^2_{P^*}=\avg{P^*}(1-\avg{P^*})$.  The
mean protein level also has the same form for both input signals,
$\avg{X}=\rho\avg{P^*}/\mu$. However, differences in the timing of protein
production will mean that the variance in the protein level differs between
the two signals. For a constant signal the variance
\begin{equation*} \label{eq:X_var_osc}
	\sigma^2_{X, {\rm c}}=\avg{X}_{\rm c}\left[1+
		\frac{\rho \lambda}{(\alpha\kappa+\lambda)(\alpha\kappa+\lambda+\mu)}
	\right]
\end{equation*}
consists of an intrinsic Poissonian term due to randomness in protein
production, and an extrinsic contribution from fluctuations in the promoter
state. The variance given an oscillatory input $\sigma^2_{X,{\rm o}}$ can
similarly be derived, but the full expression is unwieldy and thus not
presented here. In the following we will compare $\sigma^2_{X,{\rm o}}$ with
$\sigma^2_{X,{\rm c}}$ at the same mean response level, achieved by choosing
the level of the constant signal $\alpha=\alpha(\tau,T,\kappa,\lambda)$ such
that $\avg{P^*}_{\rm c}=\avg{P^*}_{\rm o}=\avg{P^*}$, and hence also
$\avg{X}_{\rm c}=\avg{X}_{\rm o}=\avg{X}$.

It is instructive to first consider cases in which the two input signals
lead to similar distributions for $X$. First, if $\rho,\mu\rightarrow\infty$
with $X_\infty=\rho/\mu$ held constant, whenever the promoter is inactive
the protein level is $X(t)=0$, while when the promoter is active the protein
level will be Poisson-distributed with mean $X_\infty$. The variance in $X$
is then given by $\sigma_{X,{\rm c}}^2=\sigma_{X,{\rm
o}}^2=\avg{X}[1+X_\infty-\avg{X}]$. In this limit of fast protein dynamics
the precise pattern of promoter switching does not affect the variance in
protein expression. At the other extreme, if either promoter switching or
protein production is slow compared to the oscillating input
($\kappa,\lambda\ll T^{-1}$ or $\rho,\mu\ll T^{-1}$), the slow reactions
effectively integrate over the temporal variation of the input. Since the
network dynamics is too slow to reliably respond to the oscillating signal,
the protein response is equivalent to that for the constant input.

\begin{figure}
	\includegraphics{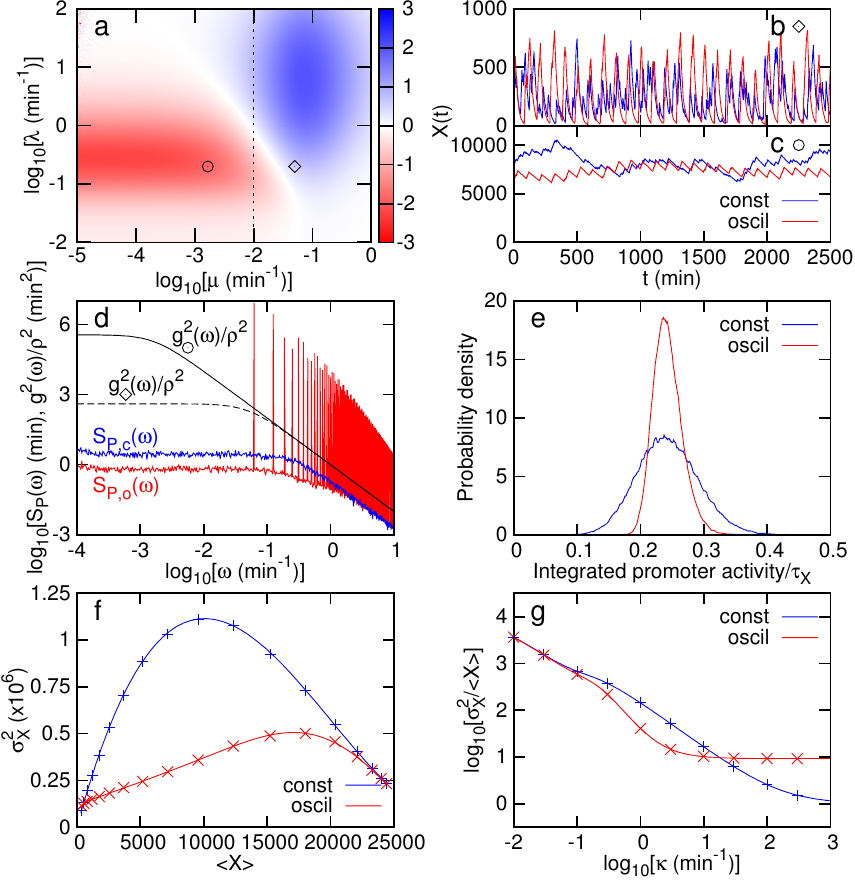}
	\caption{	\label{fig:promoter}
		(a) Relative noise for oscillating and constant signals,
		$\log_2(\sigma^2_{X,{\rm o}}/\sigma^2_{X,{\rm c}})$, as the protein
		degradation rate $\mu$ and promoter deactivation rate $\lambda$ are varied.
		In red (blue) regions the oscillatory (constant) signal leads to lower 
		noise. Other parameters are: $\tau=25\m$, $T=100\m$, $\kappa=1\m^{-1}$,
		$\rho=50\m^{-1}$. The dotted line indicates $T^{-1}$.
		(b,c) Typical time-series of $X(t)$ for the parameter combinations denoted 
		in (a) by $\diamond$ (b: $\lambda=0.2\m^{-1},\mu=0.05\m^{-1}$) and $\circ$ 
		(c: $\lambda=0.2\m^{-1},\mu=0.00167\m^{-1}$). Unless otherwise noted, 
		parameters as in (c) are used in panels (d-g).
		(d) The promoter power spectrum $S_{P}(\omega)$ and gain
		$g^2(\omega)/\rho^2=[\omega^2+\mu^2]^{-1}$. For both inputs there is a
		noise background due to randomness in promoter switching. Peaks appear
		in $S_{P,{\rm o}}(\omega)$ at $\omega=n\omega_T$ due to the periodicity
		of the oscillating input signal.
		(e) Distribution of the fraction of time the promoter is active in an
		interval $\tau_{\rm X}$, calculated from stochastic simulations. The
		oscillatory signal leads to more reproducible promoter activity on the
		timescale $\tau_{\rm X}$.
		(f) Variance against mean output level as $T$ is varied with $\tau$ held
		constant. The oscillating signal can achieve a lower variance over nearly 
		the full range of output levels. In (f) and (g), lines show exact analytic
		results, points show results of stochastic simulations.
		(g) Fano factor $\sigma^2_{X}/\avg{X}$ as $\kappa$ is varied with constant
		$K=\lambda/\kappa=0.2$. For slow switching $\kappa\tau\lesssim1$ the two
		signals become equivalent. For extremely fast switching the constant signal
		minimizes $\sigma^2_X$. At intermediate switching rates, 
		$0.1\m^{-1}\lesssim\kappa\lesssim18\m^{-1}$, the oscillating signal allows 
		for smaller $\sigma^2_X$.
	}
\end{figure}

Differences appear between the variances for the oscillatory and constant
signals in the biologically most important regime of intermediate parameter
values, where the promoter is able to respond to the oscillating input
signal, protein production can react to switching of the promoter, and the
protein lifetime is sufficiently long that patterns of promoter activity are
important. Figures~\ref{fig:promoter}(a-c) show that there exist regions in
which either the constant or oscillating signal leads to smaller
fluctuations $\sigma^2_X$ over large parameter ranges. Here we consider
parameter ranges representative of eukaryotic cells, in which promoter
switching occurs on a timescale of minutes to hours \cite{Chubb06,Suter11}
and protein lifetimes are in the range of a few to hundreds of hours
\cite{Schwanhausser11}. The timescale of the input signal is chosen to be
representative of NF-$\kappa$B oscillations \cite{Nelson04}. However, we
emphasise that out results are general and would apply equally to
prokaryotic cells in an appropriate parameter regime.

To understand the noise properties for the two input signals we consider the
network response in the frequency domain. The dynamics of $X(t)$ can be
described by the Langevin equation
\begin{equation} \label{eq:X_langevin}
	\diff{X}{t}=\rho P^*(t)-\mu X(t)+\eta(t),
\end{equation}
where $\eta(t)$ represents Gaussian white noise with $\avg{\eta(t)}=0$ and
$\avg{\eta(t)\eta(t')}=
\left(\rho\mathbb{E}[P^*(t)]+\mu\mathbb{E}[X(t)]\right)\delta(t-t')$
\cite{Warren06}. Using the spectral addition rule \cite{TanaseNicola06} it
is straightforward to calculate from Eq.~\ref{eq:X_langevin} the power
spectrum of fluctuations in $X(t)$,
$S_X(\omega)=
\left[\rho^2S_{P}(\omega)+S_{\eta}(\omega)\right]/(\omega^2+\mu^2)$, in
terms of the spectra of promoter fluctuations $S_P(\omega)$ and of intrinsic
noise in the production and decay of $X$, $S_\eta(\omega)$.  The variance
$\sigma^2_X$ can be found by integrating $S_X(\omega)$ over all frequencies,
and can therefore be written as the sum of intrinsic and extrinsic terms,
$\sigma^2_X=\sigma^2_{\rm ex}+\sigma^2_{\rm in}$ with
\begin{equation} \label{eq:noise_def}
	\sigma^2_{\rm ex}=\int_{0}^{\infty}g^2(\omega)S_{P}(\omega)d\omega,\ \ 
	\sigma^2_{\rm in}=
		\int_{0}^{\infty}\frac{S_{\eta}(\omega)}{\omega^2+\mu^2}d\omega,
\end{equation}
where $g^2(\omega)=\rho^2/(\omega^2+\mu^2)$ is a signal-independent gain
factor. The intrinsic noise $\sigma^2_{\rm in}=\avg{X}$ is also independent
of the input signal. Any differences between $\sigma_{X,{\rm o}}^2$ and
$\sigma_{X,{\rm c}}^2$ must, therefore, arise from differences in
$S_P(\omega)$.

For a constant input, $S_{P,{\rm c}}(\omega)=
2\sigma^2_{P^*}\tau_P/(\omega^2\tau_P^2+1)$ (Fig.~\ref{fig:promoter}(d), 
blue line) has a simple Lorentzian form due to random promoter switching,
with $\tau_P=(\alpha\kappa+\lambda)^{-1}$ the switching correlation time. A
general expression for $S_{P,{\rm o}}(\omega)$ with an oscillating input is
more difficult to calculate. However, simulations show
(Fig.~\ref{fig:promoter}(d), red line) that there are two significant
components. First, there are sharp peaks at frequencies corresponding to the
signal period $\omega_T=2\pi/T$ and multiples thereof, reflecting systematic
changes in $\mathbb{E}[P^*(t)]$ due to the periodicity of $s(t)$. Second,
$S_{P,{\rm o}}(\omega)$ includes an approximately Lorentzian noise
background associated with random (Poissonian) switching of the promoter
when $s(t)=1$ and delays in deactivation when $s(t)=0$. 

The protein lifetime $\tau_{\rm X}=\mu^{-1}$ relative to the signal period
$T$ is particularly important in determining which signal minimizes the
output noise.  The constant signal typically leads to smaller fluctuations
when $\tau_{\rm X}<T$ (see Fig.~\ref{fig:promoter}(a)). In this regime, for
the oscillatory signal $\sigma_{\rm ex}^2$ is dominated by contributions
from the peaks of $S_{P,{\rm o}}(\omega)$ appearing at $\omega=n\omega_T$,
since $\mu\gtrsim\omega_T$ and the gain $g^2(\omega)$ is large for
frequencies $\omega<\mu$ (Fig.~\ref{fig:promoter}(d), dashed line).
Consequently, $X(t)$ features large production bursts when the signal is
``on'' with most proteins decaying before the next input pulse, while for a
constant signal production and decay are more regularly distributed (see
Fig.~\ref{fig:promoter}(b)).

For long protein lifetimes $\tau_{\rm X}>T$, $\sigma_{X,{\rm o}}^2$ is
typically smaller than $\sigma_{X,{\rm c}}^2$ (see
Fig.~\ref{fig:promoter}(a)).  In this regime the impact of production bursts
is reduced because proteins produced during many previous signal periods
contribute to $X(t)$. Since $\mu<\omega_T$, the region $\omega\lesssim\mu$
where $g^2(\omega)$ is largest does not reach the first peak of $S_{P,{\rm
o}}(\omega)$ at $\omega_T$ (Fig.~\ref{fig:promoter}(d), solid black line);
hence $\sigma^2_{\rm ex}$ is dominated by promoter switching noise at low
frequencies $\omega<\omega_T$, for which $S_{P,{\rm o}}(\omega)<S_{P,{\rm
c}}(\omega)$. The large amplitude changes of the oscillatory signal strongly
bias the promoter to be active during a signal pulse and to be inactive
between pulses, which in turn greatly reduces the probability of observing
very long periods of promoter (in)activity. The elimination of such slow
promoter fluctuations, which lead to the largest fluctuations in $X(t)$,
means that on the timescale of the protein lifetime, promoter activity
becomes more reproducible and the production of proteins becomes less
variable when driven by an oscillatory signal (Fig.~\ref{fig:promoter}(e)).
Furthermore, Fig.~\ref{fig:promoter}(f) shows that in this regime
$\sigma_{X,{\rm o}}^2 \leq\sigma_{X,{\rm c}}^2$ over nearly the full range
of expression levels as the signal period $T$ is varied, indicating that
this result does not require fine tuning of the reaction rates to the
oscillation timescale.

Output noise $\sigma^2_X$ tends to decrease as promoter switching is made
faster (Fig.~\ref{fig:promoter}(g)): increasing the promoter switching rate
reduces $S_P(\omega)$ at low frequencies, shifting power instead to high
frequencies where $g^2(\omega)\sim\omega^{-2}$ is small. Interestingly, for
extremely fast switching the constant signal is able to achieve a smaller
variance. In this limit noise in promoter switching becomes negligible, and
Eq.~\ref{eq:promoter} reduces to 
\begin{equation*} \label{eq:birth-death}
	\emptyset\xrightarrow{\rho's(t)}\mol{X} \xrightarrow{\mu}\emptyset.
\end{equation*}
With a constant signal the effective production rate of $\mol{X}$ is
$\rho'_{\rm c}=\rho\avg{P^*}/\alpha$ and $\sigma_{X,{\rm c}}^2=\avg{X}=
\sigma_{\rm in}^2$; $\sigma_{\rm ex}^2$ (Eq.~\ref{eq:noise_def}) can be made
arbitrarily small by shifting all promoter fluctuations to extremely high
frequencies. With an oscillatory input the effective production rate becomes
$\rho'_{\rm o}=\rho/(1+K)$, where $K=\lambda/\kappa$. The resulting
variance,
\begin{equation*} \label{eq:bd_fm_noise}
	\sigma_{X,{\rm o}}^2=\avg{X}\bigg[1+\frac{\rho'_o}{\mu}
		\bigg(1-\frac{\tau}{T}-
		\frac{(1-e^{-\mu\tau})(1-e^{-\mu(T-\tau)})}{\mu\tau(1-e^{-\mu T})}
		\bigg)
	\bigg],
\end{equation*}
includes a (positive) extrinsic contribution. While the promoter switching
noise background in $S_{P,{\rm o}}(\omega)$ vanishes, as with a constant
signal, the peaks due to the periodicity of $s(t)$ remain. Hence for an
oscillatory input there will always be some overlap between $S_{P,{\rm o}}
(\omega)$ and $g^2(\omega)$, and a non-zero extrinsic noise $\sigma^2_{\rm
ex}$. With the reaction rates representative of eukaryotic transcription
used in Fig.~\ref{fig:promoter}(g), the cross-over at which $\sigma_{X,{\rm
o}}^2=\sigma_{X,{\rm c}}^2$ is $\kappa\approx18\m^{-1}$.
Experimentally-determined rates of promoter activation, however, are
typically $0.01-1\m^{-1}$ \cite{Chubb06,Suter11}, which suggests that under
biologically-relevant conditions oscillatory signals can lead to more robust
protein expression.

Thus far we have assumed that the signals $s(t)$ are deterministic. In
reality, these signals will also be noisy as they are themselves generated
by stochastic biochemical processes. An important question is whether
oscillatory signals can still be decoded more reliably once noise in the
input stimulus is taken into account. 

We first consider the effect of transcription factor copy-number
fluctuations around a constant mean by simulating $s(t)$ as a birth-death
process. We find that $\sigma_X^2$ at fixed $\avg{X}$ increases
monotonically as the variance or correlation time of $s(t)$ are increased.
Therefore, without additional non-exponential temporal correlations in the
input signal, noise is unable to reduce $\sigma_X^2$ below that achieved
with $s(t)=\alpha$.

\begin{figure}
	\includegraphics{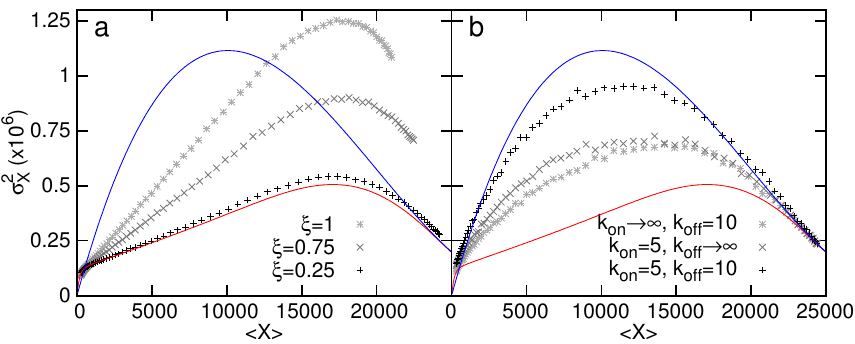}
	\caption{ \label{fig:noise}
		Effects of noise in (a) oscillation amplitude, and (b) the duration of 
		``on'' and ``off'' periods. Solid lines show results for constant (blue)
		and deterministic oscillatory (red) stimuli. Even with significant 
		variability in amplitude or timing, the oscillating signal leads to a 
		smaller $\sigma_X^2$ than a constant signal. Parameters are as in 
		Fig.~\ref{fig:promoter}(f).
		}
\end{figure}

For an oscillatory input signal there are two principal types of noise:
fluctuations in the amplitude and timing of signal pulses. First we perform
simulations in which on each occasion that the oscillatory signal switches
``on'' the amplitude $s(t)=a$ is sampled from a log-normal distribution with
mean $\overline{a}=1$ and width parameterized by
$\xi^2=\ln[\overline{a^2}]$. Figure \ref{fig:noise}(a) shows that
$\sigma_X^2$ is largely unchanged until the noise in the pulse amplitude
becomes large. Even with $\xi=1$ ($\sigma_a/\overline{a}\approx1.3$) the
output noise is typically smaller than for a constant, noiseless, input
signal. Intuitively, in the regime where an oscillating signal leads to a
smaller $\sigma^2_X$, $K\ll1$; the promoter activity is driven to saturation
during a signal pulse and hence amplitude fluctuations have little effect
until there is a significant probability that $a\sim K$. The decoding of
large-amplitude oscillatory signals is therefore highly robust to noise in
the oscillation amplitude.

We next investigate by simulations the effect of noise in the duration of
signal pulses and inter-pulse intervals. The duration of each ``on'' period
is chosen independently from a gamma distribution with mean $\tau$ and shape
parameter $k_{\rm on}$. The resulting variance in ``on'' durations is
$\sigma^2_{t_{\rm on}}=\tau^2/k_{\rm on}$. ``Off'' periods are similarly
sampled from a distribution with mean $T-\tau$ and parameter $k_{\rm off}$.
Figure~\ref{fig:noise}(b) shows that the noisy oscillation can lead to
protein level fluctuations which are similar to or smaller than
$\sigma^2_{X,{\rm c}}$ even when variability in signal timing is significant
($k_{\rm on}=5$ gives $\sigma_{t_{\rm on}}/\tau\approx0.45$).
Experimentally-observed fluctuations in oscillation periods or peak widths
vary between different systems but can be $20-30\%$
\cite{GevaZatorsky06,Shankaran09}, suggesting that in vivo oscillations can
be sufficiently precise as to reduce output noise compared to a
constant-amplitude signal.

The simple model of gene expression considered here neglects mRNA dynamics
and processing. Such processes can affect the propagation of promoter-state
fluctuations in two ways. First, mRNA dynamics will integrate over promoter
fluctuations on timescales shorter than the mRNA lifetime (typically tens to
hundreds of minutes in eukaryotic cells \cite{Rabani11,Schwanhausser11}).
However, differences between $\sigma^2_{X,{\rm c}}$ and $\sigma^2_{X,{\rm
o}}$ are primarily due to promoter-state fluctuations on timescales
comparable to or longer than the slowest timescale of variations in the
protein level, which is typically determined by the protein lifetime (ten to
a hundred hours \cite{Schwanhausser11}). Promoter fluctuations on these long
timescales can not be filtered by the mRNA dynamics, and hence even with
mRNA dynamics taken into account an oscillatory input will lead to more
robust expression since an oscillatory signal suppresses promoter
fluctuations on timescales longer than the signal period (see
Fig.~\ref{fig:promoter}(d)).  Second, a (random) delay between transcription
initiation and protein synthesis will be introduced. However, such delays
will only significantly affect the number of proteins produced on the
timescale $\tau_{\rm X}$ if the width of the delay distribution itself
becomes comparable to the protein lifetime, which seems unrealistic.  Hence
we conclude that mRNA dynamics will have little effect on our results.

It is believed that biochemical networks employ frequency-encoding schemes
in which stimuli are represented in the frequency of oscillations of
signaling molecules \cite{Berridge88,Cai08}. Our results suggest that
frequency-encoding may allow for more reliable signaling than
amplitude-encoding schemes because oscillatory signals can be decoded more
reliably. 

\begin{acknowledgments}
	We thank Andrew Mugler for comments on the manuscript. This work is part
	of the research program of the ``Stichting voor Fundamenteel Onderzoek der
	Materie (FOM)'', which is financially supported by the ``Nederlandse
	organisatie voor Wetenschappelijk Onderzoek (NWO)''.
\end{acknowledgments}


\begin{thebibliography}{10}

\bibitem{Behar10}
M.~Behar and A.~Hoffmann.
\newblock {\em Curr. Opin. Genet. Dev.}, 20:684, 2010.

\bibitem{Paszek10}
P.~Paszek, D.A. Jackson, and M.R.H. White.
\newblock {\em Curr. Opin. Genet. Devel.}, 20:670, 2010.

\bibitem{Berridge88}
M.J. Berridge and A.~Galione.
\newblock {\em FASEB J.}, 2:3074, 1988.

\bibitem{Shankaran09}
H.~Shankaran {\em et al}.
\newblock {\em Mol. Sys. Biol.}, 5:332, 2009.

\bibitem{Nelson04}
D.E.~Nelson {\em et al}.
\newblock {\em Science}, 306:704, 2004.

\bibitem{Jacquet03}
M.~Jacquet {\em et al}.
\newblock {\em J. Cell Biol.}, 161:497, 2003.

\bibitem{Cai08}
L.~Cai, C.K. Dalal, and M.B. Elowitz.
\newblock {\em Nature}, 455:485, 2008.

\bibitem{Lahav04}
G.~Lahav {\em et al}.
\newblock {\em Nat. Genet.}, 36:147, 2004.

\bibitem{Trump95}
B.F. Trump and I.K. Berezesky.
\newblock {\em FASEB J.}, 9:219, 1995.

\bibitem{Gall00}
D.~Gall, E.~Baus, and G.~Dupont.
\newblock {\em J. Theor. Biol.}, 207:445, 2000.

\bibitem{Rapp81}
P.E. Rapp, A.I. Mees, and C.T. Sparrow.
\newblock {\em J. Theor. Biol.}, 90:531, 1981.

\bibitem{Suter11}
D.M.~Suter {\em et al}.
\newblock {\em Science}, 322:472, 2011.

\bibitem{VanKampen}
N.G. {van Kampen}.
\newblock {\em {Stochastic Processes in Physics and Chemistry}}.
\newblock North Holland, Amsterdam, 1992.

\bibitem{Gillespie76}
D.T. Gillespie.
\newblock {\em J. Comput. Phys.}, 22:403, 1976.

\bibitem{Chubb06}
J.R. Chubb {\em et al}.
\newblock {\em Curr. Biol.}, 16:1018, 2006.

\bibitem{Schwanhausser11}
B.~Schwanh\"{a}usser {\em et al}.
\newblock {\em Nature (London)}, 473:337, 2011.

\bibitem{Warren06}
P.B. Warren, S.~T\u{a}nase-Nicola, and P.R. {ten Wolde}.
\newblock {\em J. Chem. Phys.}, 125:144904, 2006.

\bibitem{TanaseNicola06}
S.~T\u{a}nase-Nicola, P.B. Warren, and P.R. {ten Wolde}.
\newblock {\em Phys. Rev. Lett.}, 97:068102, 2006.

\bibitem{GevaZatorsky06}
N.~Geva-Zatorsky {\em et al}.
\newblock {\em Mol. Syst. Biol.}, 2:2006.0033, 2006.

\bibitem{Rabani11}
M.~Rabani {\em et al}.
\newblock {\em Nat. Biotechnol.}, 29:436, 2011.

\end{thebibliography}
\end{document}